\documentclass[aps,pra,twocolumn,showpacs,preprintnumbers,amsmath,amssymb,superscriptaddress,10pt]{revtex4-1}

\usepackage{amsmath,amssymb}
\usepackage[latin1]{inputenc}
\usepackage{ae}

\begin{document}
\title{Comment on ``Resilience of gated avalanche photodiodes against bright illumination attacks in quantum cryptography''}

\author{Lars Lydersen}
\email{lars.lydersen@iet.ntnu.no}
\author{Vadim Makarov}
\author{Johannes Skaar}
\affiliation{Department of Electronics and Telecommunications, Norwegian University of Science and Technology, NO-7491 Trondheim, Norway}
\affiliation{University Graduate Center, NO-2027 Kjeller, Norway}

\maketitle
\thispagestyle{empty}

Quantum key distribution (QKD) has initially been proven secure using ideal devices. However, implementations use imperfect devices available with current technology. Therefore, there are security proofs for QKD which model the devices to allow these imperfection, though at the expense of a lower secure key rate. To achieve provable security, it is crucial that the devices and implementations are verified to be within the models in the security proofs.

Security loopholes have been found originating from discrepancies between the actual implementations and the models in the security proofs. For instance, one such discrepancy allows the tailored bright illumination attacks \cite{lydersen2010a,lydersen2010b,wiechers2011}, recently shown also to be applicable against superconducting single-photon detectors \cite{lydersen2011b,lydersen2011c}. In this case the loophole is caused by the response of qubit measurement devices (detectors) to swarms of qubits (bright illumination). The question is how to counter such loopholes.

In their paper, Yuan \textit{et al.}\ propose to counter these bright illumination attacks by monitoring the avalanche photodiode (APD) current for ``anomalously high values'' \cite{yuan2011}. The robustness of this countermeasure is shown by arguing that previously proposed attacks do not work anymore. First of all, this leaves the challenge of determining what is ``anomalously high''. In order to achieve provable security, this threshold must originate from a security proof. Secondly, the fundamental issue, namely that the detector response deviates from the models in the security proofs \footnote{There are security proofs including this detector response in their model of the receiver (for instance Ref.~\cite{maroy2010}), but they predict zero secret key rate for such receivers.}, is not solved by this countermeasure.

As discussed previously \cite{yuan2010,lydersen2010c}, practical QKD cannot become provably secure by intuitive countermeasures against known attacks. This approach also requires manufacturers to make frequent, possibly costly upgrades to their systems. Loopholes should instead be countered by modifying the implementation and/or the security proofs such that the devices are within the models of the security proofs. This is the only way practical QKD can obtain the provable security that makes it superior to classical key distribution schemes. This is also how loopholes have been handled previously:
for example, the photon-number splitting attack \cite{brassard2000} led to more general security proofs \cite{gottesman2004} and eventually more efficient protocols to negate the decrease in the key rate \cite{hwang2003}.
In another example, detector efficiency mismatch \cite{makarov2006}, enabling for instance the time-shift attack \cite{qi2007,zhao2008}, is now included in security proofs \cite{fung2009,maroy2010}. 
For the bright illumination attacks, we have proposed a secure detection scheme which integrates with security proofs \cite{lydersen2011a}. In this scheme, a calibrated light source is used to verify the quantum efficiency in the center of the detector gate. Randomizing detection events outside the center of the gate provides a lower bound on the fraction of detections in the center of the gate. 

In this particular case, we have already shown that an eavesdropper using temporally tailored light of short pulses containing less than 120 photons can threaten the security of QKD \cite{lydersen2011b}. This faint after-gate attack would not be detectable with the countermeasure proposed by Yuan \textit{et al.}, since the pulses would not cause an ``anomalously high'' current, but rather a current similar to the current caused by a single photon. Therefore, this serves as an example of the risk associated with closing loopholes in an intuitive way.


\begin{thebibliography}{18}%
\makeatletter
\providecommand \@ifxundefined [1]{%
 \@ifx{#1\undefined}
}%
\providecommand \@ifnum [1]{%
 \ifnum #1\expandafter \@firstoftwo
 \else \expandafter \@secondoftwo
 \fi
}%
\providecommand \@ifx [1]{%
 \ifx #1\expandafter \@firstoftwo
 \else \expandafter \@secondoftwo
 \fi
}%
\providecommand \natexlab [1]{#1}%
\providecommand \enquote  [1]{``#1''}%
\providecommand \bibnamefont  [1]{#1}%
\providecommand \bibfnamefont [1]{#1}%
\providecommand \citenamefont [1]{#1}%
\providecommand \href@noop [0]{\@secondoftwo}%
\providecommand \href [0]{\begingroup \@sanitize@url \@href}%
\providecommand \@href[1]{\@@startlink{#1}\@@href}%
\providecommand \@@href[1]{\endgroup#1\@@endlink}%
\providecommand \@sanitize@url [0]{\catcode `\\12\catcode `\$12\catcode
  `\&12\catcode `\#12\catcode `\^12\catcode `\_12\catcode `\%12\relax}%
\providecommand \@@startlink[1]{}%
\providecommand \@@endlink[0]{}%
\providecommand \url  [0]{\begingroup\@sanitize@url \@url }%
\providecommand \@url [1]{\endgroup\@href {#1}{\urlprefix }}%
\providecommand \urlprefix  [0]{URL }%
\providecommand \Eprint [0]{\href }%
\providecommand \doibase [0]{http://dx.doi.org/}%
\providecommand \selectlanguage [0]{\@gobble}%
\providecommand \bibinfo  [0]{\@secondoftwo}%
\providecommand \bibfield  [0]{\@secondoftwo}%
\providecommand \translation [1]{[#1]}%
\providecommand \BibitemOpen [0]{}%
\providecommand \bibitemStop [0]{}%
\providecommand \bibitemNoStop [0]{.\EOS\space}%
\providecommand \EOS [0]{\spacefactor3000\relax}%
\providecommand \BibitemShut  [1]{\csname bibitem#1\endcsname}%
\let\auto@bib@innerbib\@empty
\bibitem [{\citenamefont {Lydersen}\ \emph
  {et~al.}(2010{\natexlab{a}})\citenamefont {Lydersen}, \citenamefont
  {Wiechers}, \citenamefont {Wittmann}, \citenamefont {Elser}, \citenamefont
  {Skaar},\ and\ \citenamefont {Makarov}}]{lydersen2010a}%
  \BibitemOpen
  \bibfield  {author} {\bibinfo {author} {\bibfnamefont {L.}~\bibnamefont
  {Lydersen}}, \bibinfo {author} {\bibfnamefont {C.}~\bibnamefont {Wiechers}},
  \bibinfo {author} {\bibfnamefont {C.}~\bibnamefont {Wittmann}}, \bibinfo
  {author} {\bibfnamefont {D.}~\bibnamefont {Elser}}, \bibinfo {author}
  {\bibfnamefont {J.}~\bibnamefont {Skaar}}, \ and\ \bibinfo {author}
  {\bibfnamefont {V.}~\bibnamefont {Makarov}},\ }\href@noop {} {\bibfield
  {journal} {\bibinfo  {journal} {Nat. Photonics}\ }\textbf {\bibinfo {volume}
  {4}},\ \bibinfo {pages} {686} (\bibinfo {year}
  {2010}{\natexlab{a}})}\BibitemShut {NoStop}%
\bibitem [{\citenamefont {Lydersen}\ \emph
  {et~al.}(2010{\natexlab{b}})\citenamefont {Lydersen}, \citenamefont
  {Wiechers}, \citenamefont {Wittmann}, \citenamefont {Elser}, \citenamefont
  {Skaar},\ and\ \citenamefont {Makarov}}]{lydersen2010b}%
  \BibitemOpen
  \bibfield  {author} {\bibinfo {author} {\bibfnamefont {L.}~\bibnamefont
  {Lydersen}}, \bibinfo {author} {\bibfnamefont {C.}~\bibnamefont {Wiechers}},
  \bibinfo {author} {\bibfnamefont {C.}~\bibnamefont {Wittmann}}, \bibinfo
  {author} {\bibfnamefont {D.}~\bibnamefont {Elser}}, \bibinfo {author}
  {\bibfnamefont {J.}~\bibnamefont {Skaar}}, \ and\ \bibinfo {author}
  {\bibfnamefont {V.}~\bibnamefont {Makarov}},\ }\href@noop {} {\bibfield
  {journal} {\bibinfo  {journal} {Opt. Express}\ }\textbf {\bibinfo {volume}
  {18}},\ \bibinfo {pages} {27938} (\bibinfo {year}
  {2010}{\natexlab{b}})}\BibitemShut {NoStop}%
\bibitem [{\citenamefont {Wiechers}\ \emph {et~al.}(2011)\citenamefont
  {Wiechers}, \citenamefont {Lydersen}, \citenamefont {Wittmann}, \citenamefont
  {Elser}, \citenamefont {Skaar}, \citenamefont {Marquardt}, \citenamefont
  {Makarov},\ and\ \citenamefont {Leuchs}}]{wiechers2011}%
  \BibitemOpen
  \bibfield  {author} {\bibinfo {author} {\bibfnamefont {C.}~\bibnamefont
  {Wiechers}}, \bibinfo {author} {\bibfnamefont {L.}~\bibnamefont {Lydersen}},
  \bibinfo {author} {\bibfnamefont {C.}~\bibnamefont {Wittmann}}, \bibinfo
  {author} {\bibfnamefont {D.}~\bibnamefont {Elser}}, \bibinfo {author}
  {\bibfnamefont {J.}~\bibnamefont {Skaar}}, \bibinfo {author} {\bibfnamefont
  {C.}~\bibnamefont {Marquardt}}, \bibinfo {author} {\bibfnamefont
  {V.}~\bibnamefont {Makarov}}, \ and\ \bibinfo {author} {\bibfnamefont
  {G.}~\bibnamefont {Leuchs}},\ }\href@noop {} {\bibfield  {journal} {\bibinfo
  {journal} {New J. Phys.}\ }\textbf {\bibinfo {volume} {13}},\ \bibinfo
  {pages} {013043} (\bibinfo {year} {2011})}\BibitemShut {NoStop}%
\bibitem [{\citenamefont {Lydersen}\ \emph
  {et~al.}({\natexlab{a}})\citenamefont {Lydersen}, \citenamefont {Jain},
  \citenamefont {Wittmann}, \citenamefont {Mar{\o}y}, \citenamefont {Skaar},
  \citenamefont {Marquardt}, \citenamefont {Makarov},\ and\ \citenamefont
  {Leuchs}}]{lydersen2011b}%
  \BibitemOpen
  \bibfield  {author} {\bibinfo {author} {\bibfnamefont {L.}~\bibnamefont
  {Lydersen}}, \bibinfo {author} {\bibfnamefont {N.}~\bibnamefont {Jain}},
  \bibinfo {author} {\bibfnamefont {C.}~\bibnamefont {Wittmann}}, \bibinfo
  {author} {\bibfnamefont {{\O}.}~\bibnamefont {Mar{\o}y}}, \bibinfo {author}
  {\bibfnamefont {J.}~\bibnamefont {Skaar}}, \bibinfo {author} {\bibfnamefont
  {C.}~\bibnamefont {Marquardt}}, \bibinfo {author} {\bibfnamefont
  {V.}~\bibnamefont {Makarov}}, \ and\ \bibinfo {author} {\bibfnamefont
  {G.}~\bibnamefont {Leuchs}},\ }\Eprint
  {http://arxiv.org/abs/1106.2119} {arXiv:1106.2119 [quant-ph]} \BibitemShut
  {NoStop}%
\bibitem [{\citenamefont {Lydersen}\ \emph
  {et~al.}({\natexlab{b}})\citenamefont {Lydersen}, \citenamefont {Akhlaghi},
  \citenamefont {Majedi}, \citenamefont {Skaar},\ and\ \citenamefont
  {Makarov}}]{lydersen2011c}%
  \BibitemOpen
  \bibfield  {author} {\bibinfo {author} {\bibfnamefont {L.}~\bibnamefont
  {Lydersen}}, \bibinfo {author} {\bibfnamefont {M.~K.}\ \bibnamefont
  {Akhlaghi}}, \bibinfo {author} {\bibfnamefont {A.~H.}\ \bibnamefont
  {Majedi}}, \bibinfo {author} {\bibfnamefont {J.}~\bibnamefont {Skaar}}, \
  and\ \bibinfo {author} {\bibfnamefont {V.}~\bibnamefont {Makarov}},\
  }\Eprint
  {http://arxiv.org/abs/1106.2396} {arXiv:1106.2396 [quant-ph]} \BibitemShut
  {NoStop}%
\bibitem [{\citenamefont {Yuan}\ \emph {et~al.}(2011)\citenamefont {Yuan},
  \citenamefont {Dynes},\ and\ \citenamefont {Shields}}]{yuan2011}%
  \BibitemOpen
  \bibfield  {author} {\bibinfo {author} {\bibfnamefont {Z.~L.}\ \bibnamefont
  {Yuan}}, \bibinfo {author} {\bibfnamefont {J.~F.}\ \bibnamefont {Dynes}}, \
  and\ \bibinfo {author} {\bibfnamefont {A.~J.}\ \bibnamefont {Shields}},\
  }\href@noop {} {\bibfield  {journal} {\bibinfo  {journal} {Appl. Phys.
  Lett.}\ }\textbf {\bibinfo {volume} {98}},\ \bibinfo {eid} {231104} (\bibinfo
  {year} {2011})}\BibitemShut {NoStop}%
\bibitem [{Note1()}]{Note1}%
  \BibitemOpen
  \bibinfo {note} {There are security proofs including this detector response
  in their model of the receiver (for instance Ref.~\cite {maroy2010}), but
  they predict zero secret key rate for such receivers.}\BibitemShut {Stop}%
\bibitem [{\citenamefont {Yuan}\ \emph {et~al.}(2010)\citenamefont {Yuan},
  \citenamefont {Dynes},\ and\ \citenamefont {Shields}}]{yuan2010}%
  \BibitemOpen
  \bibfield  {author} {\bibinfo {author} {\bibfnamefont {Z.~L.}\ \bibnamefont
  {Yuan}}, \bibinfo {author} {\bibfnamefont {J.~F.}\ \bibnamefont {Dynes}}, \
  and\ \bibinfo {author} {\bibfnamefont {A.~J.}\ \bibnamefont {Shields}},\
  }\href@noop {} {\bibfield  {journal} {\bibinfo  {journal} {Nat. Photonics}\
  }\textbf {\bibinfo {volume} {4}},\ \bibinfo {pages} {800} (\bibinfo {year}
  {2010})}\BibitemShut {NoStop}%
\bibitem [{\citenamefont {Lydersen}\ \emph
  {et~al.}(2010{\natexlab{c}})\citenamefont {Lydersen}, \citenamefont
  {Wiechers}, \citenamefont {Wittmann}, \citenamefont {Elser}, \citenamefont
  {Skaar},\ and\ \citenamefont {Makarov}}]{lydersen2010c}%
  \BibitemOpen
  \bibfield  {author} {\bibinfo {author} {\bibfnamefont {L.}~\bibnamefont
  {Lydersen}}, \bibinfo {author} {\bibfnamefont {C.}~\bibnamefont {Wiechers}},
  \bibinfo {author} {\bibfnamefont {C.}~\bibnamefont {Wittmann}}, \bibinfo
  {author} {\bibfnamefont {D.}~\bibnamefont {Elser}}, \bibinfo {author}
  {\bibfnamefont {J.}~\bibnamefont {Skaar}}, \ and\ \bibinfo {author}
  {\bibfnamefont {V.}~\bibnamefont {Makarov}},\ }\href@noop {} {\bibfield
  {journal} {\bibinfo  {journal} {Nat. Photonics}\ }\textbf {\bibinfo {volume}
  {4}},\ \bibinfo {pages} {801} (\bibinfo {year}
  {2010}{\natexlab{c}})}\BibitemShut {NoStop}%
\bibitem [{\citenamefont {Brassard}\ \emph {et~al.}(2000)\citenamefont
  {Brassard}, \citenamefont {L\"utkenhaus}, \citenamefont {Mor},\ and\
  \citenamefont {Sanders}}]{brassard2000}%
  \BibitemOpen
  \bibfield  {author} {\bibinfo {author} {\bibfnamefont {G.}~\bibnamefont
  {Brassard}}, \bibinfo {author} {\bibfnamefont {N.}~\bibnamefont
  {L\"utkenhaus}}, \bibinfo {author} {\bibfnamefont {T.}~\bibnamefont {Mor}}, \
  and\ \bibinfo {author} {\bibfnamefont {B.~C.}\ \bibnamefont {Sanders}},\
  }\href@noop {} {\bibfield  {journal} {\bibinfo  {journal} {Phys. Rev. Lett.}\
  }\textbf {\bibinfo {volume} {85}},\ \bibinfo {pages} {1330} (\bibinfo {year}
  {2000})}\BibitemShut {NoStop}%
\bibitem [{\citenamefont {Gottesman}\ \emph {et~al.}(2004)\citenamefont
  {Gottesman}, \citenamefont {Lo}, \citenamefont {L{\" u}tkenhaus},\ and\
  \citenamefont {Preskill}}]{gottesman2004}%
  \BibitemOpen
  \bibfield  {author} {\bibinfo {author} {\bibfnamefont {D.}~\bibnamefont
  {Gottesman}}, \bibinfo {author} {\bibfnamefont {H.-K.}\ \bibnamefont {Lo}},
  \bibinfo {author} {\bibfnamefont {N.}~\bibnamefont {L{\" u}tkenhaus}}, \ and\
  \bibinfo {author} {\bibfnamefont {J.}~\bibnamefont {Preskill}},\ }\href@noop
  {} {\bibfield  {journal} {\bibinfo  {journal} {Quant. Inf. Comp.}\ }\textbf
  {\bibinfo {volume} {4}},\ \bibinfo {pages} {325} (\bibinfo {year}
  {2004})}\BibitemShut {NoStop}%
\bibitem [{\citenamefont {Hwang}(2003)}]{hwang2003}%
  \BibitemOpen
  \bibfield  {author} {\bibinfo {author} {\bibfnamefont {W.-Y.}\ \bibnamefont
  {Hwang}},\ }\href@noop {} {\bibfield  {journal} {\bibinfo  {journal} {Phys.
  Rev. Lett.}\ }\textbf {\bibinfo {volume} {91}},\ \bibinfo {pages} {057901}
  (\bibinfo {year} {2003})}\BibitemShut {NoStop}%
\bibitem [{\citenamefont {Makarov}\ \emph {et~al.}(2006)\citenamefont
  {Makarov}, \citenamefont {Anisimov},\ and\ \citenamefont
  {Skaar}}]{makarov2006}%
  \BibitemOpen
  \bibfield  {author} {\bibinfo {author} {\bibfnamefont {V.}~\bibnamefont
  {Makarov}}, \bibinfo {author} {\bibfnamefont {A.}~\bibnamefont {Anisimov}}, \
  and\ \bibinfo {author} {\bibfnamefont {J.}~\bibnamefont {Skaar}},\
  }\href@noop {} {\bibfield  {journal} {\bibinfo  {journal} {Phys. Rev. A}\
  }\textbf {\bibinfo {volume} {74}},\ \bibinfo {pages} {022313} (\bibinfo
  {year} {2006})};\ \bibinfo {note} {Erratum ibid. \textbf{78}, 019905
  (2008)}\BibitemShut {NoStop}%
\bibitem [{\citenamefont {Qi}\ \emph {et~al.}(2007)\citenamefont {Qi},
  \citenamefont {Fung}, \citenamefont {Lo},\ and\ \citenamefont {Ma}}]{qi2007}%
  \BibitemOpen
  \bibfield  {author} {\bibinfo {author} {\bibfnamefont {B.}~\bibnamefont
  {Qi}}, \bibinfo {author} {\bibfnamefont {C.-H.~F.}\ \bibnamefont {Fung}},
  \bibinfo {author} {\bibfnamefont {H.-K.}\ \bibnamefont {Lo}}, \ and\ \bibinfo
  {author} {\bibfnamefont {X.}~\bibnamefont {Ma}},\ }\href@noop {} {\bibfield
  {journal} {\bibinfo  {journal} {Quant. Inf. Comp.}\ }\textbf {\bibinfo
  {volume} {7}},\ \bibinfo {pages} {73} (\bibinfo {year} {2007})}\BibitemShut
  {NoStop}%
\bibitem [{\citenamefont {Zhao}\ \emph {et~al.}(2008)\citenamefont {Zhao},
  \citenamefont {Fung}, \citenamefont {Qi}, \citenamefont {Chen},\ and\
  \citenamefont {Lo}}]{zhao2008}%
  \BibitemOpen
  \bibfield  {author} {\bibinfo {author} {\bibfnamefont {Y.}~\bibnamefont
  {Zhao}}, \bibinfo {author} {\bibfnamefont {C.-H.~F.}\ \bibnamefont {Fung}},
  \bibinfo {author} {\bibfnamefont {B.}~\bibnamefont {Qi}}, \bibinfo {author}
  {\bibfnamefont {C.}~\bibnamefont {Chen}}, \ and\ \bibinfo {author}
  {\bibfnamefont {H.-K.}\ \bibnamefont {Lo}},\ }\href@noop {} {\bibfield
  {journal} {\bibinfo  {journal} {Phys. Rev. A}\ }\textbf {\bibinfo {volume}
  {78}},\ \bibinfo {eid} {042333} (\bibinfo {year} {2008})}\BibitemShut
  {NoStop}%
\bibitem [{\citenamefont {Fung}\ \emph {et~al.}(2009)\citenamefont {Fung},
  \citenamefont {Tamaki}, \citenamefont {Qi}, \citenamefont {Lo},\ and\
  \citenamefont {Ma}}]{fung2009}%
  \BibitemOpen
  \bibfield  {author} {\bibinfo {author} {\bibfnamefont {C.-H.~F.}\
  \bibnamefont {Fung}}, \bibinfo {author} {\bibfnamefont {K.}~\bibnamefont
  {Tamaki}}, \bibinfo {author} {\bibfnamefont {B.}~\bibnamefont {Qi}}, \bibinfo
  {author} {\bibfnamefont {H.-K.}\ \bibnamefont {Lo}}, \ and\ \bibinfo {author}
  {\bibfnamefont {X.}~\bibnamefont {Ma}},\ }\href@noop {} {\bibfield  {journal}
  {\bibinfo  {journal} {Quant. Inf. Comp.}\ }\textbf {\bibinfo {volume} {9}},\
  \bibinfo {pages} {131} (\bibinfo {year} {2009})}\BibitemShut {NoStop}%
\bibitem [{\citenamefont {Mar{\o}y}\ \emph {et~al.}(2010)\citenamefont
  {Mar{\o}y}, \citenamefont {Lydersen},\ and\ \citenamefont
  {Skaar}}]{maroy2010}%
  \BibitemOpen
  \bibfield  {author} {\bibinfo {author} {\bibfnamefont {{\O}.}~\bibnamefont
  {Mar{\o}y}}, \bibinfo {author} {\bibfnamefont {L.}~\bibnamefont {Lydersen}},
  \ and\ \bibinfo {author} {\bibfnamefont {J.}~\bibnamefont {Skaar}},\
  }\href@noop {} {\bibfield  {journal} {\bibinfo  {journal} {Phys. Rev. A}\
  }\textbf {\bibinfo {volume} {82}},\ \bibinfo {pages} {032337} (\bibinfo
  {year} {2010})}\BibitemShut {NoStop}%
\bibitem [{\citenamefont {Lydersen}\ \emph {et~al.}(2011)\citenamefont
  {Lydersen}, \citenamefont {Makarov},\ and\ \citenamefont
  {Skaar}}]{lydersen2011a}%
  \BibitemOpen
  \bibfield  {author} {\bibinfo {author} {\bibfnamefont {L.}~\bibnamefont
  {Lydersen}}, \bibinfo {author} {\bibfnamefont {V.}~\bibnamefont {Makarov}}, \
  and\ \bibinfo {author} {\bibfnamefont {J.}~\bibnamefont {Skaar}},\
  }\href@noop {} {\bibfield  {journal} {\bibinfo  {journal} {Phys. Rev. A}\
  }\textbf {\bibinfo {volume} {83}},\ \bibinfo {pages} {032306} (\bibinfo
  {year} {2011})}\BibitemShut {NoStop}%
\end{thebibliography}

%
\end{document}